\documentstyle[12pt]{article}
\begin{document}
\begin{center}

{\bf The model of particle production by strong external sources}\\

I.M. Dremin\footnote{Email: dremin@lpi.ru}\\

Lebedev Physical Institute RAS, 119991 Moscow, Russia\\

\end{center}
\begin{abstract}
Using some knowledge of multiplicity disributions for high energy reactions,
it is possible to propose a simple analytical model of particle production 
by strong external sources. The model describes qualitatively most peculiar
properties of the distributions. The generating function of the distribution 
varies so drastically as it can happen at phase transitions.
\end{abstract}


\section{Introduction}

Multiparticle production is the main outcome of high energy collisions.
Quantum chromodynamics (QCD) provides the general framework for its 
description. It is especially successful in predicting the properties of 
hard processes. The perturbative approach with some additional phenomenological
assumptions is effective there (for reviews see \cite{dr, koch, dgar}).
The non-perturbative methods are waited for because the bulk of produced partons
have a small fraction $x$ of the longitudinal momentum of the incoming partners.
It was proposed \cite{mcl1, mcl2, mcl3} to describe them as a classical color
field rather than as particles. The large $x$ partons are then considered as
color sources for the classical field. Many QCD studies of this idea have 
been done (for recent review see \cite{leon}).

To avoid technical complications of gauge theories, it is proposed \cite{gve1, gve2}
to consider the simpler theory of a real scalar field $\phi $ with strong time-dependent
external sources $j$. The Lagrangian density of the theory is
\begin{equation}
{\cal L}\equiv \frac {1}{2}\partial _{\mu }\phi \partial ^{\mu } \phi - 
\frac {1}{2}m^2\phi ^2-\frac {g}{3!}\phi ^3 + j\phi.    \label{lagr}
\end{equation}
It represents the toy model of the Color Glass Condensate formalism. One hopes that its 
qualitative features could also be valid for particle production in QCD. Moreover, one
can use the special behavior of multiplicity distributions, gained from QCD studies and
confirmed by experiment, to develop the more detailed model in the framework of such a
theory. This is the main goal of the present paper.

\section{General relations}

The multiplicity distribution is the most general feature of multiparticle production 
processes. Any other inclusive (e.g., rapidity) distribution is obtained by averaging
over different multiplicities. To be successful, any phenomenological model of these
processes should, first of all, fit the experimental values for probabilities of
$n$-particle production $P_n$ or, equivalently, the moments of this distribution.
The latter ones are easily computed with the generating function
\begin{equation}
G(u)=\sum _{n=0}^{\infty }P_nu^n.   \label{gu}
\end{equation}
The unnormalized factorial moments are defined as
\begin{equation}
{\cal F}_q=\frac {d^qG(u)}{du^q}\vert _{u=1}=\sum _{n=0}^{\infty }n(n-1)...(n-q+1)P_n,  \label{fq}
\end{equation}
and the unnormalized cumulant moments are
\begin{equation}
{\cal K}_q=\frac {d^q\ln G(u)}{du^q}\vert _{u=1}.   \label{kq}
\end{equation}
They are connected by the iterative relations
\begin{equation}
{\cal F}_q=\sum _{m=o}^{q-1}\frac {(q-1)!}{m!(q-m-1)!}{\cal K}_{q-m}{\cal F}_m.    \label{fqkq}
\end{equation}
The mean of the distribution of multiplicities is
\begin{equation}
\langle n \rangle = \sum nP_n\equiv {\cal F}_1\equiv {\cal K}_1.   \label{nav}
\end{equation}
Factorial moments are always positive. Let us remind that for the Poisson distribution
the normalized factorial moments $F_q={\cal F}_q/\langle n\rangle ^q$ are identically equal
to 1, and the normalized cumulant moments $K_q={\cal K}_q/\langle n\rangle ^q$ are equal to
0 except of $K_1=1$. For NBD, both factorial and cumulant moments are positive.

It has been shown \cite{gve1, gve2} that in the theory defined by the Lagrangian
density (\ref{lagr}) the probability to produce $n$ particles to all orders in the
coupling $g$ reads
\begin{equation}
P_n={\rm e}^{-a/g^2}\sum _{p=1}^{n}\frac {1}{p!}\sum _{\alpha _1+...+\alpha _p=n}
\frac {b_{\alpha _1}...b_{\alpha _p}}{g^{2p}}.     \label{pn}
\end{equation}
Here $a=\sum _{r=1}^{\infty }b_r$, $p$ is the number of disconnected Feynman
subdiagrams of the theory producing the $n$ particles, and $b_r/g^2$ denotes the 
contribution of all $r$-particle cuts through the connected vacuum-vacuum diagrams.

The generating function reads
\begin{equation}
G(u)=\exp [\frac {1}{g^2}\sum _{r=1}^{\infty }b_r(u^r-1)].    \label{gum}
\end{equation}

The average multiplicity is
\begin{equation}
\langle n\rangle = \frac {1}{g^2}\sum _{r=1}^{\infty }rb_r.    \label{navm}
\end{equation}

The cumulant moments are
\begin{equation}
{\cal K}_q=\frac {1}{g^2}\sum _{r=1}^{\infty }r(r-1)...(r-q+1)b_r.    \label{kqm}
\end{equation}

Cumulant moments of $P_n$ play the role of factorial moments of $b_r/g$ 
(compare (\ref{kqm}) and (\ref{fq})). The first non-zero terms of the sum
begin with $b_q$.
 
The "cut" procedure analogous to the Cutkosky rules \cite{cutk} has been
defined \cite{gve1} for the diagrams of the theory  with Lagrangian (\ref{lagr}).
The probability of having $p$ cut subdiagrams is equal to
\begin{equation}
{\cal R}_p=\frac {1}{p!}\left (\frac {a}{g^2}\right )^p{\rm e}^{-a/g^2}.     \label{rp}
\end{equation}
The average number of cut subgraphs is determined as
\begin{equation}
\langle n_{cut}\rangle=\frac {a}{g^2}.        \label{ncut}
\end{equation}

\section{The model}

In principle, the values of $b_r$ can be computed \cite{gve1, gve2} with the help
of Schwinger-Keldysh formalism \cite{schw, keld} or, at leading order, in terms of
a pair of solutions of the classical equations of motion. At present, this has not 
been done yet explicitly even though all general expressions are given in \cite{gve1, gve2}. 
Surely, some simplifications will be necessary to solve this formidable field theory problem.

In this situation, one can advocate the phenomenological approach and try to guess
such behavior of $b_r$ which would mimic the shape of the multiplicity distribution
favored by QCD and experimental data. If successful, this attempt can indicate 
general features of $b_r$ and help in approximate computing them from the set of
Feynman diagrams.

The most astonishing regularity of behavior of multiplicity distributions predicted
by QCD \cite{dr1, dne1, dne2} and confirmed by experiment (see review \cite{dgar}) is the
oscillations of the cumulant moments as functions of their rank $q$. According to
(\ref{kqm}), this peculiar shape can be only obtained if $b_r$ change their sign with 
$r$. To get negative values of $b_r$, the phases of cut diagrams in Schwinger-Keldysh 
approach should play important role. At least for some $r$,  $b_r$ can not be 
interpreted as an "intrinsic" $r$-particle production probability as done in
\cite{gve1}. Correspondingly, the relation to the reggeon calculus and AGK 
cancellations becomes not obvious. At the same time, this oscillation should 
not lead to unphysical negative values of $\langle n\rangle $ in (\ref{navm}). 

The two-parameter ($x, \phi $) model with oscillating $b_r$, we propose to use, is
\begin{equation}
b_r=g^2\frac {x^r}{r!}\cos \phi (r-1),     \label{mod}
\end{equation}
where
\begin{equation}
x=\frac {b_1}{g^2}=P_1{\rm e} ^{a/g^2}>0.    \label{xdef}
\end{equation}
The generating function and its logarithm read
\begin{equation}
G(u)=\exp[{\rm e} ^{xu\cos \phi }\cos(xu\sin \phi -\phi)-{\rm e} ^{x\cos \phi }\cos(x\sin \phi - \phi )],   \label{gee}
\end{equation}
\begin{equation}
\ln G(u)={\rm e} ^{xu\cos \phi }\cos(xu\sin \phi -\phi)-{\rm e} ^{x\cos \phi }\cos(x\sin \phi - \phi ).    \label{lngm}
\end{equation}
Note the extremely strong and rapidly varying dependence of $G(u)$ on $u$. Its exponent is 
an exponential function itself multiplied by the oscillating factor!

The cumulant moments are given by
\begin{equation}
{\cal K}_q=x^q{\rm e} ^{x\cos \phi }\cos (x\sin \phi + (q-1)\phi ).      \label{kqos}
\end{equation}
They increase exponentially with $q$ and oscillate equidistantly in the $q$-plane\footnote{For NBD,
they increase even faster (as $q!$) and do not oscillate being always positive.}. Their zeros
$q_m^{(0)}$ are positioned at
\begin{equation}
q_m^{(0)}=1+\frac {\pi (2m-1)}{2\phi }-\frac {x\sin \phi }{\phi }.     \label{kqze}
\end{equation}
($m$ is an integer number). The distance between the neighboring zeros is defined only by the 
parameter $\phi $
\begin{equation}
\Delta q^{(0)}=\frac {\pi }{\phi }.         \label{zedi}
\end{equation}
The average multiplicity is
\begin{equation}
\langle n\rangle \equiv {\cal K}_1= x{\rm e} ^{x\cos \phi }\cos (x\sin \phi ).      \label{muco}
\end{equation}
The necessary requirement of positive $\langle n\rangle $ asks for $x\sin \phi $ to be
positioned in the hemispheres where $\cos (x\sin \phi )$ is positive. Moreover, from the 
requirement $q_1^{(0)}>1$ one obtains for $\phi >0$
\begin{equation}
0<x\sin \phi <\pi /2.        \label{xsin}
\end{equation}
The probability of having $p$ cut subdiagrams (\ref{rp}) is determined by  
\begin{equation}
\frac {a}{g^2}={\rm e} ^{x\cos \phi }\cos (x\sin \phi -\phi ) - \cos \phi . \label{ag2}
\end{equation}
 
Since both $\langle n\rangle $ and $q_m^{(0)}$ depend on energy, the parameters $x$
and $\phi $ are also functions of energy. However, the product $x\sin \phi $ must be
constant because any its dependence would induce quite strange "quasi-oscillating"
component in energy behavior of the mean multiplicity. Both $\langle n\rangle $ and $q_m^{(0)}$
should increase with energy tending asymptotically to infinity. For $q_m^{(0)}$, this property
follows from QCD asymptotics of $K_q$ where the leading order results survive and all cumulant
moments become positive. Then asymptotically $\phi $ tends to 0 and $x$ tends to infinity
so that the product $x\phi $ stays (or tends to) constant. For $\langle n\rangle $ and 
$q_1^{(0)}$ one obtains 
\begin{equation}
\langle n\rangle = x{\rm e} ^x \cos (x\phi ),      \label{nas}
\end{equation}
\begin{equation}
q_1^{(0)} = 1+\frac {\pi-2x\phi }{\phi }.     \label{q0as}
\end{equation}
If $x\phi \rightarrow 0$, then the asymptotics of $\langle n\rangle $ and $q_1^{(0)}$
are defined by the parameters $x$ and $\phi $ correspondingly and, therefore, completely
independent.

The QCD-like results for the energy dependence of $\langle n\rangle $ can be obtained with
$x \propto \sqrt {\ln s}$ at $s \rightarrow \infty $ (up to NLO corrections). For 
$x \propto \ln s$, one gets the power increase of the mean multiplicity with energy
typical for hydrodynamics \cite{land} and fixed coupling QCD \cite{dhw1, dhw2}.

Insering ansatz (\ref{mod}) in Eq. (\ref{pn}) one gets the general form of the
multiplicity distribution in terms of $x$ and $\phi $
\begin{equation}
P_n=\frac {x^n}{n!}f_n(\phi ),    \label{pnxp}
\end{equation}
where
\begin{equation}
f_n(\phi )=\sum _{k=0}^{n-1}c_{n,k}\cos ^k\phi .   \label{fng}
\end{equation}
Unfortunately, no analytical expression for the coefficients $c_{n,k}$ is 
obtained. For lowest multiplicities, they are $c_{2,0}=1, c_{2,1}=1;
c_{3,0}=0, c_{3,1}=3, c_{3,2}=2; c_{4,0}=2, c_{4,1}=1, c_{4,2}=11, c_{4,3}=4.$

The model is completely fixed if its parameters $x$ and $\phi $ are replaced by the average
multiplicity and the position of the first zero of cumulant moments. By the order of magnitude,
these values are similar in different reactions at high energies \cite{dnbs, sgsa}. To be
more definitive, let us use our experience in $e^+e^-$-collisions at LEP energies \cite{dgar}. 
There, the first zero of cumulant moments $q_1^{(0)}$
is located near the value $q=4$. The average total multiplicity is about 40. These values
can be approximately obtained with the model parameters
\begin{equation}
x=3, \;\;\;\;\; \phi = \pi /12.      \label{parm}
\end{equation}
The only inconsistency with experiment comes from the distance between zeros of cumulant
moments which equals 12 that is much larger than the experimentally observed ones.
However, it would be very naive to expect even this quite exotic quantity to be
perfectly reproduced by such a simple phenomenological model. Further refinements
of the expression (\ref{mod}) are possible to get better agreement (e.g., the nonlinear
dependence of the cosine argument on $r$).

Another problem appears for hadronic reactions if the same parameters are used. Particle
production is often interpreted in terms of reggeon exchanges. In \cite{gve1} it is
proposed to identify cut disconnected vacuum-vacuum diagrams as cut reggeons. Then
the average number of these diagrams (\ref{ncut}) for the above values of $x$ and $\phi $
is very large $\langle n_{cut}\rangle \approx 14$ compared to the typical values in 
widely used reggeon models, and the average number of particles in a single
cut subgraph is quite small $\langle n\rangle _1\approx 3$. Therefore, the simple 
identification of cut subdiagrams with cut reggeons looks rather improbable.

One of the most important features of the model is the violent behavior of the 
generating function in the complex $u$-plane. Its extrema are positioned at
\begin{equation}
u_k=\frac {\pi (2k-1)}{2x\sin \phi }     \label{guex}
\end{equation}
($k$ is an integer number) with maxima and minima replacing one another at each
subsequent $k$ and separated by the distance
\begin{equation}
\Delta u=\frac {\pi }{x\sin \phi }.        \label{delu}
\end{equation}
It is especially interesting that the first extremum $u_1$ is positioned at $u_1=2$
for the above chosen values of the parameters $x$ and $\phi $. The generating function
reaches extremely high value at this point
\begin{equation}
G(u_1)\equiv G(2)\approx \exp [\frac {\pi }{12}{\rm e} ^6] \gg 1.    \label{gu1}
\end{equation}
At the next minimum $u_2=6$ the generating function is very small
\begin{equation}
G(u_2)\equiv G(6)\approx \exp [-\frac {\pi}{12}{\rm e}^{18}] \ll 1.        \label{gu2}
\end{equation}
Maxima and minima are separated by $\Delta u = 4$.
 
In some way the generating function reminds the grand canonical function of the
statistical mechanics with $z=u-1$ interpreted as fugacity. It was shown 
\cite{lyan, ylee} that the singularity of the grand canonical function at
$u=2$ corresponds to the phase transition in the condensed matter. The obtained
large value of $G(2)$ indicates that the model probably describes the processes
with drastically varying characteristics. The point $u=2$ is the accumulating 
point of zeros of the cut generating functions
\begin{equation}
G_c(u)=\sum _{n=1}^{n_{max}}P_nu^n          \label{gcut}
\end{equation}
for $n_{max}\rightarrow \infty $. These zeros are located \cite{dr2, dewo} near
the circle $\vert u-1\vert = 1$ in the complex $u$-plane. Actually, only the
cut generating functions are available from experiment because of finite $n_{max}$
measurable at finite energies. It is the polynomial of $u$ normalized at the
center of the circle $\vert u-1\vert = 1$ by the condition $G_c(1)=1$ with zero
at $u=0$ and other $n_{max}-1$ complex conjugated zeros near this circle. For
example, for $G_c(n_{max}=3)$ these two zeros are at
\begin{equation}
u_{2,3}=-\frac {P_2}{2P_3} \pm i\frac {\sqrt {4P_3P_1-P_2^2}}{2P_3} \label{u23}
\end{equation}
with
\begin{equation}
\vert u_{2,3}\vert ^2=\frac {P_1}{P_3} \;\; (\ll 1), \;\;\;\; 
{\rm Re}u_{2,3}=-\frac {P_2}{2P_3} <0.                      \label{mu23}
\end{equation}
At larger $n_{max}$, ${\rm Re} U_{n_{max}}$ increases and tends to 2 with
${\rm Im} u_{n_{max}}$ coming closer to the circle $\vert u-1\vert=1$.
It is remarkable that accumulation of zeros is related to the very high
value of $G(2)$ which reminds the singular behavior at phase transitions.

\section{Conclusions}

The simple analytical model of particle production by strong external sources
is proposed. All the moments of multiplicity distribution and its generating
function are calculable analytically. It rarely happens that one gets them
altogether. This is the priviledge of simplest distributions of the probability
theory but not of QCD.

The model describes qualitatively the multiplicity distribution of produced
particles fitting its mean value and higher order moments to experimental
data and QCD findings. In particular, the oscillations of cumulant moments,
first predicted by QCD, are reproduced. The energy behavior of these 
characteristics is related to the energy dependence of the model parameters.
Asymptotic relations help reveal the tendencies in this dependence.

The generating function of the distribution sharply depends on the auxiliary
variable $u$. This behavior is reminiscent of the grand canonical function
in statistical mechanics near the phase transitions. The exponential
function whose exponent is an exponential function by itself with an
oscillating factor in front of it is an unique guest in physics studies.
It has no analogues in the well known distributions of the probability
theory either.

Even though the results rely strongly on the expression (\ref{gum}) 
obtained for $\phi ^3$-theory with strong external sources, they can be
of more general validity. The ansatz (\ref{mod}) combined with (\ref{gum})
reproduces very tiny features of multiplicity distributions learned
from QCD and experimental studies. It can help get some insight for
developing approximate methods of computing these probabilities in
the field theory approach.

\section{Acknowledgments}

This work has been supported in part by the RFBR grants 
04-02-16445-a, 04-02-16333, 06-02-17051.\\

\end{document}